\documentclass{llncs}
\usepackage{amsmath}
\usepackage{amsfonts}
\usepackage{graphics}
\usepackage{amssymb}
\usepackage{color}
\usepackage{comment}

\newcommand{\Q}{\mathcal{Q}} \newcommand{\R}{\mathcal{R}}

\newcommand{\ra}{\rightarrow}

\newcommand{\la}{\leftarrow}

\newcommand{\dd}[2]{#1_1,\ldots,#1_{#2}}             

\newcommand{\set}[1]{\{#1\}}                      

\def\qed{\hfill{\qedboxempty}      
  \ifdim\lastskip<\medskipamount \removelastskip\penalty55\medskip\fi}

\def\qedboxempty{\vbox{\hrule\hbox{\vrule\kern3pt
                 \vbox{\kern3pt\kern3pt}\kern3pt\vrule}\hrule}}

\def\qedfull{\hfill{\qedboxfull}   
  \ifdim\lastskip<\medskipamount \removelastskip\penalty55\medskip\fi}

\def\qedboxfull{\vrule height 4pt width 4pt depth 0pt}

\newcommand{\markfull}{\qedboxfull}

 \newcommand{\wrt}[0]{w.r.t.}

\newcommand{\ifdirection}{``\underline{If}''. {}}
\newcommand{\onlyifdirection}{``\underline{Only if}''. {}}

\renewcommand{\emptyset}{\varnothing}

\newcommand{\atom}[1]{\underline{#1}}
\newcommand{\tuple}[1]{\mathbf{#1}}

\newcommand{\dom}{\Gamma}
\newcommand{\freshdom}{\Gamma_N}
\newcommand{\variables}{\Gamma_V}

\newcommand{\dep}{\Sigma}
\newcommand{\tdep}{\Sigma_T}
\newcommand{\edep}{\Sigma_E}

\newcommand{\isa}[1]{\mathit{ISA}}

\newcommand{\head}[1]{\mathit{head}(#1)}
\newcommand{\body}[1]{\mathit{body}(#1)}

\newcommand{\ans}[3]{\mathit{ans}(#1,#2,#3)}

\newcommand{\ins}[1]{\mathbf{#1}}
\newcommand{\insA}{\ins{A}}
\newcommand{\insB}{\ins{B}}

\newcommand{\insX}{\ins{X}}
\newcommand{\insY}{\ins{Y}}
\newcommand{\insZ}{\ins{Z}}

\newcommand{\chase}[2]{\mathit{chase}(#1,#2)}

\newcommand{\pchase}[3]{\mathit{chase}^{#1}(#2,#3)}

\newcommand{\mods}[2]{\mathit{mods}(#1,#2)}

\renewcommand{\paragraph}[1]{\textbf{#1}}
\newenvironment{proofsk}{\textsc{Proof (sketch).}}{$\Box$\newline}

\begin{document}

\title{Deep Separability of Ontological Constraints}
\author{Andrea Cal\`{i}$^{1,3}$ \and Marco Console$^2$ \and Riccardo
  Frosini$^2$} 
\institute{$^1$Dept.~of Computer Science and Inf.~Systems, Birkbeck University of London, UK\\
  $^2$Dip.~di Informatica e Sistemistica, Universit\`a di Roma ``La
  Sapienza'', Italy\\
  $^3$Oxford-Man Institute of Quantitative Finance, University of
  Oxford,  UK\\[2mm]
  \texttt{andrea{@}dcs.bbk.ac.uk}\\
  \{\texttt{console.marco},\texttt{frosini.riccardo}\}{@}\texttt{gmail.com}}
\maketitle

\begin{abstract}
  When data schemata are enriched with expressive constraints that aim at
  representing the domain of interest, in order to answer queries one needs to
  consider the logical theory consisting of both the data and the constraints.
  Query answering in such a context is called ontological query answering.
  Commonly adopted database constraints in this field are tuple-generating
  dependencies (TGDs) and equality-generating dependencies (EGDs).  It is well
  known that their interaction leads to intractability or undecidability of
  query answering even in the case of simple subclasses.  Several conditions
  have been found to guarantee \emph{separability}, that is lack of
  interaction, between TGDs and EGDs.  Separability makes EGDs (mostly)
  irrelevant for query answering and therefore often guarantees tractability,
  \emph{as long as the theory is satisfiable}.  In this paper we review the two
  notions of separability found in the literature, as well as several syntactic
  conditions that are sufficient to prove them.  We then shed light on the
  issue of satisfiability checking, showing that under a sufficient condition
  called \emph{deep separability} it can be done by considering the TGDs only.
  We show that, fortunately, in the case of TGDs and EGDs, separability implies
  deep separability.  This result generalizes several analogous ones, proved
  \emph{ad hoc} for particular classes of constraints.  Applications include
  the class of \emph{sticky TGDs} and EGDs, for which we provide a syntactic
  separability condition which extends the analogous one for \emph{linear
    TGDs}; preliminary experiments show the feasibility of query answering in
  this case.
\end{abstract}


\newcommand{\datalogpm}{Datalog$^\pm$}

\section{Introduction}
\label{sec:introduction}

When a database $D$ is equipped with an \emph{ontology} $\dep$, that is, a
knowledge base consisting of constraints that express relevant properties of
the underlying domain, queries are not answered merely against the database
instance $D$, but against the logical theory $D \cup \dep$.  Several languages
have been proposed for ontologies, with different computational properties.
The \emph{DL-Lite} family~\cite{ACKZ09,CDLL*07,PLCD*08} has the advantage of a
low (\textsc{ac}$_0$, which is contained in \textsc{logspace}) data complexity
of conjunctive query answering and of knowledge base satisfiability.
%
The well-known Entity-Relationship (ER)~\cite{Chen76} model has recently gained
importance in ontology specification, due to the fact that it is comprehensible
to theorists and practitioners, while having good expressive power.  The
\textsf{ER}$^\pm$ family of ER-like languages~\cite{CaGP12}, in particular,
comprises several tractable (\wrt~conjunctive query answering) ontology
languages, which properly generalize the main languages of the DL-Lite family.
Another relevant, more general class of ontology languages, is the
\emph{Datalog}$^{\pm}$~family, that is, a family of rule-based languages
derived from Datalog (see, e.g., \cite{CGLM+10} and references therein) whose
rules are (function-free) Horn rules, possibly with existentially quantified
variables in the head, called \emph{tuple-generating dependencies (TGDs)},
enriched with functionality constraints in the form of
\emph{equality-generating dependencies (EGDs)}, and \emph{negative
  constraints}, a form of denial constraints.  The simplest and least
expressive Datalog$^\pm$ languages are able to properly capture the most
prominent ontology languages, including most of the DL-lite family.

In this paper we focus on the interaction between TGDs and EGDs.  A TGD is a
first-order implication that forces the existence of tuples under certain
conditions, and it is of the form $\forall \insX \forall
\insY\,\varphi(\insX,\insY)\, \rightarrow\, \exists \insZ\,\psi(\insX,\insZ)$,
where $\varphi(\insX,\insY)$ and $\psi(\insX,\insZ)$ are conjunctions of atoms
over a relational schema.  An EGD forces equality of values under certain
conditions, and it is of the form $\forall \insX \, \varphi(\insX) \ra X_i =
X_j$, where $\varphi(\insX)$ is a conjunction of atoms over a relational
schema, and $\{X_i, X_j\} \subseteq \insX$.  To answer a query $Q$ over an
instance $D$ and a set $\dep$ of TGDs and EGDs, we could in principle expand
$D$ according to $\dep$, inferring all the entailed additional knowledge, and
then evaluate $Q$ against the obtained instance\footnote{This is not a query
  answering procedure in general, as the expansion (which is called
  \emph{chase}; see Section~\ref{sec:preliminaries}) can be infinite; however,
  the chase is a a key tool for studying the query answering problem, as it
  will be clear in Section~\ref{sec:preliminaries}.}.  In doing so, unknown
values (those corresponding to the existentially-quantified variables of TGDs)
will be represented by \emph{labelled nulls}, which are a sort of placeholder.
EGDs play a role in answering as they can enforce equalities; if an equality
between two distinct constants is enforced, the process fails as the theory is
unsatisfiable.

\begin{example} \label{exa:separability} Consider the following set $\dep$ of
  TGDs and EGDs (we omit universal quantifiers to avoid clutter):
\[
\begin{array}{rlcrl}
  \sigma_1: &  r_1(X) \ra \exists Y\,r_2(X,Y) &~& \sigma_4: & r_4(X,Y) \ra r_5(X,Y)\\
  \sigma_2: & r_2(X,Y) \ra r_3(X,Y) &~& \sigma_5: & r_5(X,Y) \ra r_2(X,Y)\\
  \sigma_3: & r_3(X,Y) \ra r_4(X,Y) &~& \eta: & r_3(X,Y), r_3(X,Z) \ra Y=Z  
\end{array}
\]
Notice that $\eta$ is a key dependency, imposing that atoms in $r_3$ have
unique values on their first attribute.  Now, let us take $D = \set{r_1(a),
  r_3(a,b)}$ and the (ground) Boolean conjunctive query $Q$ defined as $q \la
r_2(a,b)$ ($q$ is a propositional predicate), which simply asks whether
$r_2(a,b)$ holds.  Let us answer $Q$ by expanding $D$ according to $\dep$.  We
first obtain $r_2(a,\zeta)$ from $\sigma_1$, where $\zeta$ is a null.  From
$\sigma_3$ we get $r_4(a,b)$.  We then add $r_3(a,\zeta)$ from $\sigma_2$; at
this point we need to apply $\eta$ and enforce $\zeta = b$, so that
$r_3(a,\zeta)$ becomes $r_3(a,b)$.  Due to $\eta$, therefore, the query has
positive answer, written $D \cup \dep \models Q$.  However, even in the absence
of $\eta$, $Q$ would be answered positively.  In fact, if we proceed with the
expansion we get $r_5(a,b)$ from $\sigma_4$ and finally $r_2(a,b)$ from
$\sigma_5$.  Therefore we would have had the same result without $\eta$.  This
can be shown to hold for every query $Q$; therefore, \emph{provided that the
  theory $D \cup \dep$ is satisfiable} (that is, the expansion does not fail),
we can answer every query, for every $D$, by considering the TGDs only.  This
property is called \emph{separability}, and it defines a form of lack of
interaction (hence the name) between EGDs and TGDs in a certain set.
\end{example}

In this paper we provide the following contributions:
\begin{enumerate} \itemsep-\parsep
\item We illustrate the two notions of separability (the old and the new one)
  found in the literature, relating them to each other, and to the several
  syntactic conditions, sufficient to separability (one of the two notions),
  that have been proposed in the literature in the last decade or so.
\item We consider the problem of checking satisfiability of a theory consisting
  of an instance and a set of TGDs and EGDs.  This can be done by answering
  suitable Boolean Conjunctive queries, and this has been shown in several
  cases.  However, the proof that this technique works has always been \emph{ad
    hoc}, depending on the particular class of TGDs and EGDs.  Here, we show
  that this technique for the satisfiability check is sound and complete if a
  semantic condition, which we call \emph{deep separability}, holds.  We show
  that, fortunately, (new) separability implies deep separability.  This way,
  we unify several known results under a single general theorem, and we also
  provide a useful tool for all separable classes that are yet to be
  discovered.
\item Finally, we mention, as an application, a syntactic condition which is
  sufficient for separability of \emph{sticky sets of TGDs}~\cite{CaGP10a} and
  EGDs, we hint at our preliminary experiment with a prototype system, and we
  discuss some open problems.
\end{enumerate}


\section{Preliminaries}
\label{sec:preliminaries}

In this section we recall some basics on databases, queries, tuple-generating
dependencies, equality-generating dependencies, and the chase procedure.

\paragraph{General.}
We define the following pairwise disjoint (infinite) sets of symbols:
\emph{(i)} a set $\dom$~of {\em constants} (constitute the ``normal'' domain of
a database), \emph{(ii)} a set $\freshdom$~of \emph{labeled nulls} (used as
placeholders for unknown values, and thus can be also seen as variables), and
\emph{(iii)} a set $\variables$ of \emph{variables} (used in queries and
dependencies).  Different constants represent different values (\emph{unique
  name assumption}), while different nulls may represent the same value.  A
lexicographic order is defined on $\dom \cup \freshdom$, such that every value
in $\freshdom$ follows all those in $\dom$.
Throughout the paper, we denote by $\insX$ the sequence of variables
$X_1,\ldots,X_k$, where $k \geqslant 0$.  Also, let $[n]$ be the set
$\{1,\ldots,n\}$, for any integer $n \geqslant 1$.

A \emph{relational schema} $\R$ (or simply \emph{schema}) is a set of
\emph{relational symbols} (or \emph{predicates}), each with its associated
arity.  A \emph{position} $r[i]$ (in a schema $\R$) is identified by a
predicate $r \in \R$ and its $i$-th argument (or attribute).  A \emph{term} $t$
is a constant, null, or variable.  An \emph{atomic formula} (or simply
\emph{atom}) has the form $r(t_{1}, \ldots, t_{n})$, where $r$ is an $n$-ary
relation, and $t_{1}, \ldots, t_{n}$ are terms.
Conjunctions of atoms are often identified with the sets of their atoms.
A \emph{database (instance)} $D$ for a schema $\R$ is a (possibly infinite) set
of atoms of the form $r(\tuple{t})$ (a.k.a.~\emph{facts}), where $r$ is an
$n$-ary predicate of $\R$, and $\tuple{t} \in (\dom \cup \freshdom)^{n}$.  We
denote as $r(D)$ the set $\{\tuple{t}~|~r(\tuple{t}) \in D\}$.

A \emph{substitution} from one set of symbols $S_{1}$ to another set of symbols
$S_{2}$ is a function $h : S_{1} \rightarrow S_{2}$ defined as follows:
\emph{(i)} $\emptyset$ is a substitution (empty substitution), \emph{(ii)} if
$h$ is a substitution, then $h \cup \{X \rightarrow Y\}$ is a substitution,
where $X \in S_{1}$ and $Y \in S_{2}$, and $h$ does not already contain some $X
\rightarrow Z$ with $Y \neq Z$.  If $X \rightarrow Y\, \in\, h$ we write $h(X) =
Y$.
A \emph{homomorphism} from a set of atoms $A_{1}$ to a set of atoms $A_{2}$,
both over the same schema $\R$, is a substitution $h: \dom \cup \freshdom \cup
\variables \rightarrow \dom \cup \freshdom \cup \variables$ such that:
\emph{(i)} if $t \in \dom$, then $h(t) = t$, and \emph{(ii)} if $r(t_{1},
\ldots, t_{n})$ is in $A_{1}$, then $h(r(t_{1}, \ldots, t_{n})) = r(h(t_{1}),
\ldots, h(t_{n}))$ is in $A_{2}$.
If there are homomorphisms from $A_1$ to $A_2$ and vice-versa, then $A_1$ and
$A_2$ are \emph{homomorphically equivalent}.  The notion of homomorphism
naturally extends to conjunctions of atoms.
Given a set of symbols $S$, two atoms $\atom{a}_1$ and $\atom{a}_2$ are
\emph{$S$-isomorphic} iff there exists a bijection $h$ such that $h(\atom{a}_1)
= \atom{a}_2$, $h^{-1}(\atom{a}_2) = \atom{a}_1$, and $h(X) = X$, for each $X
\in S$.

\paragraph{Conjunctive Queries.}
A \emph{conjunctive query (CQ)} $Q$ of arity $n$ over a schema $\R$ has the
form $q(\insX) \la \varphi(\insX,\insY)$, where $\varphi(\insX,\insY)$ is a
conjunction of atoms over $\R$, $\insX$ and $\insY$ are sequences of variables
or constants in $\dom$, and $q$ is an $n$-ary predicate that does not occur in
$\R$.   $\varphi(\insX,\insY)$ is called the \emph{body} of $q$, denoted as
$\body{q}$.  A \emph{Boolean CQ (BCQ)} is a CQ of zero arity.

The \emph{answer} to an $n$-ary CQ $Q$ of the form $q(\insX) \la
\varphi(\insX,\insY)$ over a database $D$, denoted as $Q(D)$, is the set of all
$n$-tuples $\tuple{t} \in \dom^{n}$ for which there exists a homomorphism $h :
\insX \cup \insY \rightarrow \dom \cup \freshdom$ such that
$h(\varphi(\insX,\insY)) \subseteq D$ and $h(\insX) = \tuple{t}$.
A BCQ has only the empty tuple $\langle \rangle$ as possible answer, in which
case it is said that has positive answer.  Formally, a BCQ $Q$ has
\emph{positive} answer over $D$, denoted as $D \models Q$, iff $\langle \rangle
\in Q(D)$, or, equivalently, $Q(D) \neq \emptyset$.

\paragraph{Dependencies.}
Given a schema $\R$, a \emph{tuple-generating dependency (TGD)} $\sigma$ over
$\R$ is a first-order formula $\forall \insX \forall
\insY\,\varphi(\insX,\insY) \rightarrow \exists \insZ\,\psi(\insX,\insZ)$,
where $\varphi(\insX,\insY)$ and $\psi(\insX,\insZ)$ are conjunctions of atoms
over $\R$, called the \emph{body} and the \emph{head} of $\sigma$, denoted as
$\body{\sigma}$ and $\head{\sigma}$, respectively.   Henceforth, to avoid
notational clutter, we will omit the universal quantifiers in TGDs.
Such $\sigma$ is satisfied by a database $D$ for $\R$ iff, whenever there
exists a homomorphism $h$ such that $h(\varphi(\insX,\insY)) \subseteq D$,
there exists an extension $h'$ of $h$ (i.e., $h' \supseteq h$) where
$h'(\psi(\insX,\insZ)) \subseteq D$.

\textit{Inclusion dependencies (IDs)} are the simplest class of TGDs; they have
just one body-atom and one head-atom, without repetition of variables neither
in the body nor in the head.  They are usually denoted as $r_1[\insA] \subseteq
r_2[\insB]$, where $\insA$ and $\insB$ are sets of attributes (arguments) of
$r_1$ and $r_2$ respectively.  Such an ID expresses that for each $r_1$-atom,
its values in $\insA$ have to appear in $\insB$ some $r_2$ atom.  How this is
represented by a TGD is obvious.

An \emph{equality-generating dependency (EGD)} $\eta$ over $\R$ is a
first-order formula $\forall \insX \, \varphi(\insX) \ra X_i = X_j$, where
$\varphi(\insX)$ is a conjunction of atoms over $\R$, called the \emph{body}
and denoted as $\body{\eta}$, and $X_i = X_j$ is an equality among variables of
$\insX$.  Henceforth, for brevity, we will omit the universal quantifiers in
EGDs.
Such $\eta$ is satisfied by a database $D$ for $\R$ iff, whenever there exists
a homomorphism $h$ such that $h(\varphi(\insX)) \subseteq D$, then $h(X_i) =
h(X_j)$.  Special cases of EGDs are \emph{functional dependencies (FDs)} of the
form $r: \insA \ra \insB$, where $\insA$ and $\insB$ are sets of attributes of
$r$; such a dependency is satisfied in an instance $D$ if there are no two
atoms in $D$ that have the same values on $\insA$ but different values on
$\insB$.  If the union of $\insA$ and $\insB$ is the whole set of attributes of
$r$, the FD is said to be a \emph{key dependency (KD)}.  How a FD is expressed
by EGDs is obvious.

\paragraph{CQ Answering under Dependencies.}
We now define the notion of \emph{query answering} under TGDs and EGDs.  Given a
database $D$ for $\R$, and a set $\dep$ of TGDs and EGDs over $\R$, the
\emph{models} of $D$ \wrt~$\dep$, denoted as $\mods{D}{\dep}$, is the set of
all databases $B$ such that $B \models D \cup \dep$, i.e., $B \supseteq D$ and
$B$ satisfies $\dep$.
The \emph{answer} to a CQ $Q$ \wrt~$D$ and $\dep$, denoted as
$\ans{Q}{D}{\dep}$, is the set $\{\tuple{t}~|~\tuple{t} \in Q(B),
{\rm~for~each~} B \in \mods{D}{\dep}\}$.
The \emph{answer} to a BCQ $Q$ \wrt~$D$ and $\dep$ is \emph{positive}, denoted
as $D \cup \dep \models Q$, iff $\ans{Q}{D}{\dep} \neq \emptyset$.
Note that query answering under general TGDs and EGDs is undecidable.  In fact,
this is true even in extremely simple cases such as that of IDs and
keys~\cite{ChVa85}.

We recall that the two problems of CQ and BCQ answering under TGDs and EGDs are
\textsc{logspace}-equivalent~\cite{CaGK08}.   Moreover, it is easy to see that
the query output tuple problem (as a decision version of CQ answering) and BCQ
evaluation are \textsc{ac}$_{0}$-reducible to each other.   Thus, we henceforth
focus only on the BCQ answering problem.

\paragraph{The Chase Procedure.}
The \emph{chase procedure} (or simply \emph{chase}) is a fundamental
algorithmic tool introduced for checking implication of
dependencies~\cite{MaMS79}, and later for checking query
containment~\cite{JoKl84}, also in non-relational cases~\cite{CM:EDBT2010,CM:TPLP2010}.
Informally, the chase is a process of repairing a
database \wrt~a set of dependencies so that the resulted database satisfies the
dependencies.  We shall use the term chase interchangeably for both the
procedure and its result.  The chase works on an instance through the so-called
TGD and EGD \emph{chase rules}.  The TGD chase rule comes in two different
equivalent fashions: \emph{oblivious} and \emph{restricted}~\cite{CaGK08},
where the restricted one repairs TGDs only when they are not satisfied.  In the
sequel, we focus on the oblivious one for better technical clarity.  The chase
rules follow.

\underline{\textsc{TGD Chase Rule.}} Consider a database $D$ for a schema $\R$,
and a TGD $\sigma$ of the form $\varphi(\insX,\insY) \rightarrow \exists
\insZ\,\psi(\insX,\insZ)$ over $\R$.  If $\sigma$ is {\em applicable} to $D$,
i.e., there exists a homomorphism $h$ such that $h(\varphi(\insX,\insY))
\subseteq D$, then: \emph{(i)} define $h' \supseteq h$ such that $h'(Z_{i}) =
z_{i}$, for each $Z_{i} \in \insZ$, where $z_{i} \in \freshdom$ is a ``fresh''
labeled null not introduced before, and following lexicographically all those
introduced so far, and \emph{(ii)} add to $D$ the set of atoms
$h'(\psi(\insX,\insZ))$, if not already in $D$.

\underline{\textsc{EGD Chase Rule.}} Consider a database $D$ for a schema $\R$,
and an EGD $\eta$ of the form $\varphi(\insX) \ra X_i = X_j$ over $\R$.  If
$\eta$ is \emph{applicable} to $D$, i.e., there exists a homomorphism $h$ such
that $h(\varphi(\insX)) \subseteq D$ and $h(X_i) \neq h(X_j)$, then: \emph{(i)}
if $h(X_i)$ and $h(X_j)$ are both constants of $\dom$, then there is a
\emph{hard violation} of $\eta$, and the chase \emph{fails}, otherwise
\emph{(ii)} replace each occurrence of $h(X_j)$ with $h(X_i)$, if $h(X_i)$
precedes $h(X_j)$ in the lexicographic order, or vice-versa otherwise.

Given a database $D$ and a set of dependencies $\dep = \tdep \cup \edep$, where
$\tdep$ are TGDs and $\edep$ are EGDs, the chase algorithm for $D$ and $\dep$
consists of an exhaustive application of the chase rules in a breadth-first
fashion, which leads to a (possibly infinite) database.  Roughly, the chase of
$D$ \wrt~$\dep$, denoted as $\chase{D}{\dep}$, is the (possibly infinite)
instance constructed by iteratively applying \emph{(i)} the TGD chase rule
once, and \emph{(ii)} the EGD chase rule as long as it is applicable (i.e.,
until a fixed point is reached).  A formal definition of the chase algorithm is
given, e.g., in~\cite{CaGP10a}.

\begin{example}[from~\cite{CaPi11}] \label{exa:chase} Let $\R = \{r,s\}$.
  Consider the set $\dep$ of TGDs and EGDs over $\R$ constituted by the TGDs
  $\sigma_1 = r(X,Y) \ra \exists Z \, r(Z,X),s(Z)$ and $\sigma_2 = r(X,Y) \ra
  r(Y,X)$, and the EGD $\eta = r(X,Y),r(X',Y) \ra X=X'$.  Let $D$ be the
  database for $\R$ consisting of the single atom $r(a,b)$.
  During the construction of $\chase{D}{\dep}$ we first apply $\sigma_1$, and
  we add the atoms $r(z_1,a),s(z_1)$, where $z_1$ is a ``fresh'' null of
  $\freshdom$.
  Moreover, $\sigma_2$ is applicable and we add the atom $r(b,a)$.
  Now, the EGD $\eta$ is applicable and we replace each occurrence of $z_1$
  with the constant $b$; thus, we get the atom $s(b)$.  We continue by applying
  exhaustively the chase rules as explained above.  \hfill\markfull
\end{example}

The (possibly infinite) chase of $D$ \wrt~$\dep$ is a \emph{universal model} of
$D$ \wrt~$\dep$, i.e., for each database $B \in \mods{D}{\dep}$, there exists a
homomorphism from $\chase{D}{\dep}$ to $B$ \cite{FKMP05,DeNR08}.
Using this fact it can be shown that the chase is a formal tool for query
answering under TGDs and EGDs.  In particular, given a BCQ $Q$, $D \cup \dep
\models Q$ iff $\chase{D}{\dep} \models Q$, providing that the chase does not
fail.  If the chase fails, then the set of models of $D$ w.r.t.~$\dep$ is
empty, and $D \cup \dep \models Q$ trivially.


\section{What is separability?}
\label{sec:separability}

In this section we provide a brief review of the interaction of TGDs and EGDs
in ontological query answering.

It is well know that the interaction of TGDs and EGDs leads to undecidability
of query answering; this happens even in simple sub-classes of constraints,
such as key and inclusion dependencies~\cite{Cali03,CaLR03}.  It is therefore
useful to identify classes of constraints which do not suffer from his
``harmful'' interaction.  To this aim, a key condition is that of
\emph{separability}~\cite{CaGP12,CaGP11}, which we give below.
 
\begin{definition}[Separability]\label{def:separability}
  Consider a set $\tdep$ of TGDs over a schema $\R$, and a set $\edep$ of EGDs
  over $\R$.  We say that the set $\dep = \tdep \cup \edep$ is \emph{separable}
  if, for every database $D$ for $\R$, either $\chase{D}{\dep}$ fails, or,
  $\chase{D}{\dep} \models Q$ iff $\chase{D}{\tdep} \models Q$, for every BCQ
  $Q$ over $\R$.
\end{definition}

Notably, there is another definition of separability in the literature, which
is adopted in~\cite{Cali03,CaLR03,CaGL12}.  Such a definition is similar to the
above Definition~\ref{def:separability}, but with a difference which makes it
stronger.
For the sake of completeness, we give such a definition below.

\begin{definition}[Old separability] \label{def:old-separability} Consider a
  set $\tdep$ of TGDs over a schema $\R$, and a set $\edep$ of EGDs over $\R$.
  We say that the set $\dep = \tdep \cup \edep$ is \emph{separable} if, for
  every database $D$ for $\R$, \textit{(i)} if the chase fails, then $D
  \not\models \edep$ ($D$ does not satisfy $\dep$), and \textit{(ii)} if the
  chase does not fail, we have $\chase{D}{\dep} \models Q$ iff
  $\chase{D}{\tdep} \models Q$, for every BCQ $Q$ over $\R$.
\end{definition}

The old separability is a special case of the new separability as it
enforces condition \textit{(i)}
(Definition~\ref{def:old-separability}), which we reformulate below,
calling it \emph{EGD-stability}.


\begin{definition}[EGD-stability] \label{def:egd-stable} Consider a
  set $\tdep$ of TGDs over a schema $\R$, and a set $\edep$ of EGDs
  over $\R$.  We say that the set $\dep = \tdep \cup \edep$ is
  \emph{EGD-stable} if, for every instance $D$ for $\R$, $D \models
  \edep$ implies that $\chase{D}{\dep}$ does not fail.
\end{definition}

EGD stability guarantees that the satisfiability of $D \cup \dep$ (i.e., the
existence, or non-failure, of $\chase{D}{\dep}$) can be determined by merely
checking whether $D \models \edep$.  In fact, the difference between the old
and the new definitions of separability resides in the fact that under the new
separability condition, a chase failure does not imply that the initial
database $D$ violates some EGD in $\edep$, while this holds under the old
definition.  The satisfiability check is a fundamental step in query answering
(see Section~\ref{sec:answering}), and EGD stability guarantees it can be
easily done.  However, with a more sophisticated version of the satisfiability
check (see Section~\ref{sec:satisfiability}), the old separability can be
relaxed to the new separability, giving raise to the discovery of new
(syntactic) classes that enjoy (new) separability.  Section~\ref{sec:related}
below provides an overview of the most relevant syntactic classes in the
literature.

\subsection{Related Work}
\label{sec:related}

The notion of (old) separability was first introduced in~\cite{Cali03,CaLR03},
in the context of inclusion dependencies (IDs) and key dependencies (KDs) (see
e.g.~\cite{AbHV95}).  The general idea is to define a \emph{sufficient
  syntactic condition} for separability, which can be efficiently checked.
This was done while extending an early class of IDs and KDs (see
e.g.~\cite{AbHV95}), called \emph{key-based}.  Key-based IDs and KDs were
proposed in the milestone work by Johnson and Klug~\cite{JoKl84}, and they are
in fact separable, though not defined explicitly as such.  Key-based IDs and KDs
are defined as follows\footnote{The definition in the original paper is
  different but equivalent; we choose a definition which is more clear in the
  context of this paper.}: \textit{(i)} for each relational predicate $r$,
there is \emph{only one KD} defined on it; \textit{(ii)} for each ID $r[\insX]
\subseteq s[\insY]$, where $\insX$ and $\insY$ are set of attributes of $r$ and
$s$ respectively (see~\cite{AbHV95} for the notation), \textit{(ii.a)} $\insX$
is \emph{disjoint} from any key set of attributes for $r$, and \textit{(ii.b)}
$\insY$ is \emph{properly contained} in the set of key attributes for $s$.

The more general class of \emph{non-key-conflicting (NKC) IDs}, with respect to
a set of KDs, is defined as follows: \textit{(i)} for each relational predicate
$r$, there is \emph{only one KD} defined on it; \textit{(ii)} for every ID
$r[\insX] \subseteq s[\insY]$, $\insY$ is \emph{not} a \emph{proper} superset
of the set of key attributes for $s$ (if such set exists).  The class of KDs
and NKC IDs is separable, and it properly captures the well known class of
\emph{foreign-key dependencies}.

The class of KDs and non-key conflicting IDs was generalized in~\cite{CaGL12}
to general TGDs, which are assumed to have a single atom in the head, without
loss of generality.  The idea is analogous, and the condition is as follows:
\textit{(i)} for each relational predicate $r$, there is \emph{only one KD}
defined on it; \textit{(ii)} each existentially quantified variable in the head
of a TGD must occur only once; \textit{(iii)} for each TGD $\sigma =
\varphi(\insX,\insY) \ra \exists \insZ \, r(\insX,\insZ)$, the set of
$\insX$-attributes of $r(\insX,\insZ)$ is not a proper superset of the set of
key attributes of $r$.

In~\cite{CaGP10a}, the class of non-key-conflicting TGDs was straightforwardly
extended to treat functional dependencies (FDs) rather than keys.

The literature so far discussed deals with sufficient \emph{syntactic}
conditions that guarantee, in fact, EGD-stability.  However, there are classes
of TGDs and EGDs such that EGDs are triggered in the chase, but which enjoy
separability.  

\begin{example}\label{exa:separable-not-egd-stable}
  Consider the set of TGDs and EGDs in Example~\ref{exa:separability}.  It is
  separable, but it is not EGD-stable.  In fact, if we consider the instance
  $D' = \set{r_3(a,b), r_2(a,c)}$, we have that $D' \models \edep$ but $D \cup
  \edep$ is unsatisfiable because the chase fails due to a hard violation.
\end{example}

This leads in~\cite{CaGP12} to the definition of separability as
in Definition~\ref{def:separability} in this paper.  This work deals with IDs
and KDs which express an expressive variant of the Entity-Relationship
model~\cite{Chen76}.  By means of graph-related properties, \emph{necessary and
  sufficient} syntactic conditions for separability are provided, thus defining
useful tractable classes of constraints.

The case of \emph{linear TGDs} (TGDs with a single body-atom) and KDs was later
considered in~\cite{CaPi11,CaGP11}.  In these works, sufficient syntactic
conditions are proposed that guarantee separability, without imposing
non-egd-triggerability.  The conditions are quite involved and they make use of
backward-resolution.  Interestingly, the complexity of checking the syntactic
condition, called \emph{non-conflict} condition, is the same as that of query
answering, that is \textsc{pspace}-complete in combined complexity.

At this point, we considered separability but not the problem of
satisfiability.  While separability allows us to ignore EGDs in the case of
satisfiable theories, the problem of deciding the satisfiability remains, as
well as that of determining its complexity.  This will be the subject of next
section.

\subsection{Query Answering under Separable Constraints}
\label{sec:answering}

In the case of a set $\dep = \tdep \cup \edep$ of separable TGDs and EGDs, such
that BCQ answering under $\tdep$ is decidable, given an instance $D$ and a BCQ
$Q$, to decide whether $D \cup \dep \models Q$, the following steps are needed:
\begin{enumerate} \itemsep-\parsep
\item Check whether the chase fails, that is, whether $D \cup \dep$ is
  satisfiable; if $D \cup \dep$ is unsatisfiable, then trivially $D \cup \dep
  \models Q$ (\emph{``Ex falso quodlibet''}).
\item If $D \cup \dep$ is satisfiable, then by
  Definition~\ref{def:separability} we know $\chase{D}{\dep} \models Q$ iff
  $\chase{D}{\tdep} \models Q$, therefore we check whether $\chase{D}{\tdep}
  \models Q$.
\end{enumerate}

Apart from the complexity of the satisfiability check, we have that
the complexity of query answering is the same as that of answering
under TGDs only, which is a highly desirable property.  In the case of
EGD-stable constraints, the satisfiability check amounts simply to
checking whether $D \models \edep$, which can be done in \textsc{np},
and in \textsc{ptime} (or better, in the even lower complexity class
\textsc{ac}$_0$) if we consider $\edep$ fixed.  However, in the cases
of separable but not EGD-stable constraints, the problem is to be
addressed differently; this will be the subject of
Section~\ref{sec:satisfiability}.


\section{Separability and Satisfiability}
\label{sec:satisfiability}

In this section we address the problem of deciding whether, given an instance
$D$ and a set $\dep$ of \emph{separable} TGDs and EGDs, $D \cup \dep$ is
satisfiable.  As seen in Section~\ref{sec:separability}, this preliminary check
is needed, in the case of separability, before one proceeds to answer a BCQ $Q$
by taking into account the TGDs only.

The satisfiability check is done as in~\cite{CaGL12,CaGP12}, by encoding
\emph{hard violations} of EGDs as a set $\Q_V$ of Boolean conjunctive queries.
Given a separable set $\dep = \tdep \cup \edep$, and an instance $D$, we have
that satisfiability holds if and only if, for each $Q \in \Q_V$ we have $D \cup
\tdep \not\models Q$ or, equivalently, if and only if all queries in $\Q_V$
have negative answer when evaluated against $D$ and $\tdep$ (TGDs only).  The
encoding is done as follows.  First, we need to add auxiliary facts to $D$.
For each pair of \emph{distinct} constants $c_1, c_2$ appearing in $D$ as
arguments, we add the facts $\mathit{neq}(c_1,c_2)$ and $\mathit{neq}(c_2,c_1)$
to $D$, where $\mathit{neq}/2$ is an auxiliary predicate expressing that two
constants are distinct.  Then, for each EGD $\eta$ of the form $\phi(\insX) \ra
X_i=X_j$, with $X_i$ and $X_j$ in $\insX$, we construct the BCQ $Q_\eta$ as
follows (quantifiers omitted for brevity): $Q_\eta: q \la \phi(\insX),
\mathit{neq}(X_i,X_j)$, where $q$ is a propositional predicate.  The set $\Q_V$
encoding hard violations is then defined as the set of of all $Q_\eta$ so
constructed, or equivalently $\Q_V = \bigcup_{\eta\in\edep}Q_\eta$.  It is not
difficult to prove that if $D \cup \dep \models Q_\eta$ then $\chase{D}{\dep}$
fails, and therefore $D \cup \dep$ is unsatisfiable.  However, it still remains
to determine whether $D \cup \dep \models Q_\eta$.  Notice that in the case of
non-egd-triggerable constraints we do not have this problem, as we merely need
to check whether $D \models \edep$.  In principle, separability (see
Definition~\ref{def:separability}) does not tell us anything about the cases
when the chase fails.  However, we are still in luck because we can evaluate
the (Boolean) conjunctive queries in $Q_V$ against $D \cup \tdep$ rather than
$D \cup \dep$.  This is proved, for instance, in~\cite{CaGP12,CaPi11}, but
\emph{ad hoc}, by using the properties of the particular class of constraints
involved.  Here we show a general condition that allows us to perform the
satisfiability check as above.  We first need to introduce a preliminary
condition, called \emph{deep separability}, which is apparently more
restrictive than separability (we shall then prove that it is implied by
separability).  We start by denoting by $\pchase{k}{D}{\dep}$ the result of the
first $k$ chase steps under $\dep$ applied to an instance $D$, where a step is
either an EGD or a TGD application.

\begin{definition} \label{def:deep-separability} Let $\dep$ be a set
  of TGDs and EGDs, with $\dep = \tdep \cup \edep$, where the
  dependencies in $\tdep$ are TGDs and those in $\edep$ are EGDs.  We
  say that $\dep$ is \emph{deeply separable} if, for each integer $k
  \geq 0$, for each instance $D$ with with values in $\dom \cup
  \freshdom$, and for each Boolean conjunctive query $Q$, the
  following holds: if $\pchase{k}{D}{\dep}$ exists, then \(
  \pchase{k}{D}{\dep} \models Q \textrm{~implies~} \chase{D}{\tdep}
  \models Q \).
\end{definition}

The notion of \emph{deep separability}, intuitively, guarantees that, at any
step before a possible failure, the chase does not entail any atoms that are
not entailed by the chase computed according to $\tdep$ only.
We now come to the main result about deeply separable TGDs and EGDs.  The
result states that in the case of deep separability the satisfiability check
done by answering suitable queries as above is correct and complete.

\begin{theorem} \label{the:sat-check} Let $\dep$ be a set of deeply separable
  TGDs and EGDs, with $\dep = \tdep \cup \edep$, where the dependencies in
  $\tdep$ are TGDs and those in $\edep$ are EGDs.  Let $\Q_V = \set{\dd{Q}{m}}$
  be the set of BCQs encoding violations of the EGDs in $\edep$ as from the
  above construction.  Then we have that $\chase{D}{\tdep} \models Q_i$ for
  some $i\in\set{1,\ldots,m}$ if and only if $\chase{D}{\dep}$ fails (or
  equivalently $D \cup \dep$ is unsatisfiable).
\end{theorem}

\begin{proofsk}
  We prove the two directions of the implication separately.

  \onlyifdirection By contradiction, assume $\chase{D}{\tdep} \models
  Q_i$ for some $i\in\set{1,\ldots,m}$ but $\chase{D}{\dep}$ exists.
  It is not difficult to show that if $\chase{D}{\tdep} \models Q_i$
  then also $\chase{D}{\dep} \models Q_i$; this holds because
  $\chase{D}{\tdep}$ is a universal model for $D \cup \tdep$ and
  $\chase{D}{\dep}$ is a model (not necessarily universal) for $D \cup
  \tdep$; therefore there exists a homomorphism from the former to the
  latter.
  If $\chase{D}{\dep} \models Q_i$ then the chase must necessarily
  fail.  Contradiction.

  \ifdirection Assume $\chase{D}{\dep}$ fails at step $k$ by violation of the
  EGD $\eta \in \edep$.  Then, $\pchase{k-1}{D}{\dep}$ exists; moreover, it is
  easily seen that $\pchase{k-1}{D}{\dep} \models Q_\eta$, where $Q_\eta \in
  \Q_V$ encodes the violation of $\eta$.  By the hypothesis of deep
  separability we have $\chase{D}{\tdep} \models Q_\eta$, hence the claim.
\end{proofsk}

Notice that for the ``If'' direction we do not need deep separability nor
separability; this direction of the implication holds for general TGDs and
EGDs.
Finally we show that, despite the appearances, deep separability is not a
stricted condition than separability.  In fact, separability implies deep
separability.

\begin{theorem} \label{the:sep-implies-deep-sep} Let $\dep$ be a set of TGDs
  and EGDs.  If $\dep$ is separable, then it is deeply separable.
\end{theorem}

\begin{proofsk}
  We distinguish two cases.

  \underline{Case~1: $\chase{D}{\dep}$ exists}.  In this case the claim
  obviously holds.

  \underline{Case~2: $\chase{D}{\dep}$ fails}.  Assume $\chase{D}{\dep}$ fails
  at step $k$; therefore $\pchase{k-1}{D}{\dep}$ exists.  Take $D_F$ obtained
  from $D$ by replacing each constant in $\dom$ with a fresh null from
  $\freshdom$.  Obviously $\chase{D_F}{\dep}$ exists and so does
  $\pchase{k-1}{D_F}{\dep}$.  Take any BCQ $Q$ such that
  $\pchase{k-1}{D_F}{\dep} \models Q$.  It is straightforwardly seen that
  $\chase{D_F}{\dep} \models Q$ and, due to separability, $\chase{D_F}{\tdep}
  \models Q$.  Since $\chase{D_F}{\tdep} \models Q$ is obtained from
  $\chase{D_F}{\tdep}$ by the above renaming of constants into nulls, we have
  $\chase{D}{\tdep} \models Q$.  Since this holds for every step $\ell \leq
  k-1$, the claim is proved.
\end{proofsk}

The following result immediately follows from the above.

\begin{corollary} \label{cor:answering-complexity} For every separable set
  $\dep$ of TGDs and EGDs, for every instance $D$ and for every BCQ $Q$:
  \begin{itemize} \itemsep-\parsep
  \item checking unsatisfiability of $D \cup \dep$ has the same complexity as
    query answering under $\tdep$ alone;
  \item checking whether $D \cup \dep \models Q$ has the same
    complexity as query answering under $\tdep$ alone.
  \end{itemize}
\end{corollary}

Notice that we mention \emph{unsatisfiability} rather than satisfiability
because Theorem~\ref{the:sat-check} shows a reduction from failure
(unsatisfiability) to BCQ answering under TGDs alone.


\section{Preliminary Experiments}
\label{sec:experiments}

We show here some preliminary experiments on the separability check in
ontologies.  First of all, let us describe the ontology language we have
adopted.  In~\cite{CaGP11}, a sufficient condition for separability (in the
``new'' meaning) is provided for the case of EGDs and linear TGDs (which, we
remind the reader, are TGDs with exactly one head-atom and one body-atom).  We
have extended such condition, with a rather straightforward adaptation, to the
class of \emph{sticky sets of TGDs}~\cite{CaGP10a}, a relevant class that
allows for a form of joins in the body.  The condition requires, as in the case
of~\cite{CaGP11}, a significant number of containment checks under sticky sets
of TGDs in practical cases.  More specifically, there is a first step, like
in~\cite{CaGP11}, where the body of each EGD is expanded via a (terminating)
variant of backward resolution, and then for each obtained subgoal, a suitable
containment check (under TGDs alone) is performed.  We have built a prototype
system which performs such a separability check which requires -- as is obvious
from this paper -- also the satisfiability check.  In the first tests, we
considered a few known ontologies, including the GALEN medical ontology and the
LUBM ontology (benchmark from Lehigh University, which describes the
organizational structure of universities).  
\begin{figure}[tbh]
  \centering
  \begin{tabular}{|c|c|c|}
    \hline
    \textbf{Max arity} & \textbf{Containment checks} &  \textbf{Time~(s)}\\
    \hline
    \hline  
    $4$ & 18 & 0.372 \\
    \hline
    $5$ & 61 & 1.635 \\
    \hline
    $6$ & 186 & 7.603 \\
    \hline
    $7$ & 540 & 67.797 \\
    \hline
    $8$ & 1450 & 527.608 \\
    \hline
    $9$ & 4748 & 6191.924 \\
    \hline
  \end{tabular}

  \caption{Experimental results for the separability check in "near-worst-case"
    ontologies.}
  \label{fig:experiments}
\end{figure}
The tests have been run on an Intel QuadCore~I7 at 2.0~GHz and with 4~Gb of
RAM; the operating system was Windows~7 Professional carrying Java Standard
Edition Runtime Environment 1.7.0.  The considered ontologies do not include
any expressive functionality constraints, therefore we designed some ourselves,
according to the semantics.  In such cases, separability checks ran in
extremely short time, confirming our idea that real-world functionality
constraints can be efficiently checked for separability.  We therefore moved on
to test the system on near-worst-case schemata.  We ``artificially'' designed
some ad-hoc ontologies that, to be checked for separability, require a high
number of containment checks (double exponential in the maximum predicate
arity).  Some values are shown in Figure~\ref{fig:experiments}.
Notice that execution times are still reasonable for arities that are already
too large to exist in practice.  Given that the check is independent of
queries, and therefore it is run only in case either the constraints or the
instance change, we believe the first results are promising and the
separability check is generally efficient in practice.


\section{Discussion}
\label{sec:discussion}

In this paper we have given an overview of the problem of separability between
TGDs and EGDs in the context of ontological query answering, where queries are
(Boolean) conjunctive queries.  We have reviewed the two main notions of
separability found in the literature, the ``old'' one and the ``new'' one, and
we have clarified the difference between them.  We have then addressed the
issue of checking satisfiability of a set of TGDs and EGDs together with an
instance, and we have shown that this can be done by merely answering suitable
queries in the case of deeply separable classes of constraints.  We have shown
the desirable property that all separable sets of constraints are also deeply
separable.  We have therefore clarified and proved formally a satisfiability
check which was already employed in the literature, but proved on a
case-by-case basis~\cite{CaGP11,CaGL12,CaGP12}, depending on the class of
constraints.  We believe that our generalisation provides a useful tool for
future studies, and a better insight into the satisfiability problem.
Then, we have shown some preliminary experimental results on a prototype system
we have built.  While the experimentation is still at an initial stage, the
first results are promising, even in ``pathological'' cases where we designed
ontologies that behave, with respect to the separability check, as the
(asymptotically) worst case.

Our research, given the broad variety of expressive ontology languages
captured by the formalisms we considered, has broad applications in
ontology-based data access, as we provide tools for the separability
check.  Interestingly, in Data Exchange~\cite{FKMP05}, where the TGDs
are such that the chase is guaranteed to terminate, EGDs cannot lead to undecidability of query answering.
We note that the dependencies considered in this paper may also be understood as integrity constraints whose application may be incrementally checked \cite{M:FQAS2004,CM:LOPSTR2003,CM:AAI2000}, possibly in the presence of uncertain or diversified data~\cite{DM:FlexDBIST2006,DM:FlexDBIST2007,DM:QOIS2009,FMT:SIGMOD2012}.
The results in this paper also find application in Data Exchange, a form of
ontology-based data access where rules (TGDs) ``move'' data from a source
schema to a target one, while other constraints (TGDs and EGDs) are expressed
on the target schema.

\paragraph{Open problems.}  We currently have two main open problems at hand,
which seem non-trivial.  First, we would like to study the decidability of
determining whether a given set of TGDs and EGDs is separable.  This has been
proved to be undecidable for the case of arbitrary TGDs and EGDs~\cite{Pier12};
however, the proof relies on the fact that query answering under arbitrary TGDs
(without EGDs) is undecidable~\cite{BeVa81}.  It is not clear whether
undecidability holds also in the cases where query answering is undecidable
under TGDs and EGDs together, but decidable under TGDs alone; relevant cases
are, for instance, IDs and KDs, sticky sets of TGDs and EGDs, or guarded TGDs
and KDs~\cite{CaGL09}.  We conjecture that determining separability is
undecidable for these classes, but a proof is still to be devised.  The second
problem is more technical, and is determining the complexity of checking the
syntactic condition for the separability of sticky sets of TGDs and EGDs.  We
have an obvious \textsc{exptime} lower bound, but the upper bound is currently
unknown.

\paragraph{Acknowledgments.}  We would like to thank Georg Gottlob, Andreas
Pieris and the anonymous reviewers for their insightful comments on this
research.



\bibliographystyle{abbrv}
\bibliography{amw-12}

\begin{thebibliography}{10}

\bibitem{AbHV95}
S.~Abiteboul, R.~Hull, and V.~Vianu.
\newblock {\em Foundations of Databases}.
\newblock Addison-Wesley, 1995.

\bibitem{ACKZ09}
A.~Artale, D.~Calvanese, R.~Kontchakov, and M.~Zakharyaschev.
\newblock The {DL}-{L}ite family and relations.
\newblock {\em J. Artif. Intell. Res.}, 36:1--69, 2009.

\bibitem{BeVa81}
C.~Beeri and M.~Y. Vardi.
\newblock The implication problem for data dependencies.
\newblock In {\em Proc.~of ICALP}, pages 73--85, 1981.

\bibitem{Cali03}
A.~Cal\`\i.
\newblock {\em Query Answering and Optimisation in Data Integration Systems}.
\newblock PhD thesis, Universit\`a di Roma {``La Sapienza''}, 2003.

\bibitem{CaGK08}
A.~Cal\`{\i}, G.~Gottlob, and M.~Kifer.
\newblock Taming the infinite chase: Query answering under expressive
  relational constraints.
\newblock In {\em Proc.~of KR}, pages 70--80, 2008.

\bibitem{CaGL09}
A.~Cal\`{\i}, G.~Gottlob, and T.~Lukasiewicz.
\newblock A general datalog-based framework for tractable query answering over
  ontologies.
\newblock In {\em Proc.~of PODS}, pages 77--86, 2009.

\bibitem{CaGL12}
A.~Cal\`{\i}, G.~Gottlob, and T.~Lukasiewicz.
\newblock A general datalog-based framework for tractable query answering over
  ontologies.
\newblock {\em J.~Web Sem.}, 2012.
\newblock to appear.

\bibitem{CGLM+10}
A.~Cal\`{\i}, G.~Gottlob, T.~Lukasiewicz, B.~Marnette, and A.~Pieris.
\newblock {D}atalog+/-: {A} family of logical knowledge representation and
  query languages for new applications.
\newblock In {\em Proc.~of LICS}, pages 228--242, 2010.

\bibitem{CaGP10a}
A.~Cal\`{\i}, G.~Gottlob, and A.~Pieris.
\newblock Advanced processing for ontological queries.
\newblock {\em PVLDB}, 3(1):554--565, 2010.

\bibitem{CaGP11}
A.~Cal\`{\i}, G.~Gottlob, and A.~Pieris.
\newblock Querying conceptual schemata with expressive equality constraints.
\newblock In {\em Proc.~of ER}, pages 161--174, 2011.

\bibitem{CaGP12}
A.~Cal\`{\i}, G.~Gottlob, and A.~Pieris.
\newblock Ontological query answering under expressive {E}ntity-{R}elationship
  schemata.
\newblock {\em Inf. Syst.}, 37(4):320--335, 2012.

\bibitem{CaLR03}
A.~Cal\`{\i}, D.~Lembo, and R.~Rosati.
\newblock On the decidability and complexity of query answering over
  inconsistent and incomplete databases.
\newblock In {\em Proc.~of PODS}, pages 260--271, 2003.

\bibitem{CM:TPLP2010}
A.~Cal{\`{i}} and D.~Martinenghi.
\newblock Querying incomplete data over extended er schemata.
\newblock {\em Theory and Practice of Logic Programming}, 10(3):291--329, 2010.

\bibitem{CM:EDBT2010}
A.~Cal\`{\i} and D.~Martinenghi.
\newblock Querying the deep web (tutorial).
\newblock In {\em EDBT 2010, 13th International Conference on Extending
  Database Technology, Lausanne, Switzerland, March 22-26, 2010, Proceedings},
  pages 724--727, 2010.

\bibitem{CaPi11}
A.~Cal\`{\i} and A.~Pieris.
\newblock On equality-generating dependencies in ontology querying --
  {P}reliminary report.
\newblock In {\em Proc.~of AMW}, 2011.

\bibitem{CDLL*07}
D.~Calvanese, G.~De~Giacomo, D.~Lembo, M.~Lenzerini, and R.~Rosati.
\newblock Tractable reasoning and efficient query answering in description
  logics: The {DL}-lite family.
\newblock {\em J. Autom. Reasoning}, 39(3):385--429, 2007.

\bibitem{ChVa85}
A.~K. Chandra and M.~Y. Vardi.
\newblock The implication problem for functional and inclusion dependencies.
\newblock {\em SIAM Journal of Computing}, 14:671--677, 1985.

\bibitem{Chen76}
P.~P. Chen.
\newblock The entity-relationship model: towards a unified view of data.
\newblock {\em ACM TODS}, 1(1):124--131, 1995.

\bibitem{CM:AAI2000}
H.~Christiansen and D.~Martinenghi.
\newblock {Symbolic Constraints for Meta-Logic Programming}.
\newblock {\em Applied Artificial Intelligence}, 14(4):345--367, 2000.

\bibitem{CM:LOPSTR2003}
H.~Christiansen and D.~Martinenghi.
\newblock Simplification of database integrity constraints revisited: A
  transformational approach.
\newblock In {\em Logic Based Program Synthesis and Transformation, 13th
  International Symposium LOPSTR 2003, Uppsala, Sweden, August 25-27, 2003,
  Revised Selected Papers}, volume 3018 of {\em Lecture Notes in Computer
  Science}, pages 178--197. Springer, 2004.

\bibitem{DM:FlexDBIST2006}
H.~Decker and D.~Martinenghi.
\newblock Avenues to flexible data integrity checking.
\newblock In {\em Proceedings of the International Workshop on Flexible
  Database and Information System Technology (FlexDBIST-06) 6 September 2006,
  Krakow, Poland}, pages 425--429. IEEE Computer Society, 2006.

\bibitem{DM:FlexDBIST2007}
H.~Decker and D.~Martinenghi.
\newblock Getting rid of straitjackets for flexible integrity checking.
\newblock In {\em Proceedings of the 2nd International Workshop on Flexible
  Database and Information System Technology (FlexDBIST-07)}, pages 360--364,
  2007.

\bibitem{DM:QOIS2009}
H.~Decker and D.~Martinenghi.
\newblock Modeling, measuring and monitoring the quality of information.
\newblock In {\em Proceedings of the 4th International Workshop on Quality of
  Information Systems (QoIS 2009)}, pages 212--221, 2009.

\bibitem{DeNR08}
A.~Deutsch, A.~Nash, and J.~B. Remmel.
\newblock The chase revisited.
\newblock In {\em Proc.~of PODS}, pages 149--158, 2008.

\bibitem{FKMP05}
R.~Fagin, P.~G. Kolaitis, R.~J. Miller, and L.~Popa.
\newblock Data exchange: {S}emantics and query answering.
\newblock {\em Theor. Comput. Sci.}, 336(1):89--124, 2005.

\bibitem{FMT:SIGMOD2012}
P.~Fraternali, D.~Martinenghi, and M.~Tagliasacchi.
\newblock {Top-k bounded diversification}.
\newblock In {\em Proceedings of the 2012 ACM SIGMOD/PODS Conference -- SIGMOD
  2012, Scottsdale, Arizona, USA, May 20--24, 2012}, pages 421--432, 2012.

\bibitem{JoKl84}
D.~S. Johnson and A.~C. Klug.
\newblock Testing containment of conjunctive queries under functional and
  inclusion dependencies.
\newblock {\em J. Comput. Syst. Sci.}, 28(1):167--189, 1984.

\bibitem{MaMS79}
D.~Maier, A.~O. Mendelzon, and Y.~Sagiv.
\newblock Testing implications of data dependencies.
\newblock {\em ACM Trans. Database Syst.}, 4(4):455--469, 1979.

\bibitem{M:FQAS2004}
D.~Martinenghi.
\newblock Simplification of integrity constraints with aggregates and
  arithmetic built-ins.
\newblock In {\em Flexible Query Answering Systems, 6th International
  Conference, FQAS 2004, Lyon, France, June 24-26, 2004, Proceedings}, volume
  3055 of {\em Lecture Notes in Computer Science}, pages 348--361. Springer,
  2004.

\bibitem{Pier12}
A.~Pieris, 2012.
\newblock Personal communication.

\bibitem{PLCD*08}
A.~Poggi, D.~Lembo, D.~Calvanese, G.~{De Giacomo}, M.~Lenzerini, and R.~Rosati.
\newblock Linking data to ontologies.
\newblock {\em J. Data Semantics}, 10:133--173, 2008.

\end{thebibliography}


\end{document}
